\documentclass[article,12pt]{article}
\usepackage{amsfonts}
\usepackage{graphicx}

\newcommand{\p}{\partial}
\newcommand{\n}{\nabla}

\begin{document}

\begin{center} 

\begin{LARGE} 
\begin{bf}
       A mesh adaptivity scheme on the 
       Landau-de Gennes functional minimization 
       case in 3D, and its driving efficiency \\[12pt]
\end{bf} 
\end{LARGE}

\begin{large} 
   Iztok Bajc\footnote{Fakulteta za matematiko in fiziko, Univerza v Ljubljani, Jadranska 19, Ljubljana SI-1000, Slovenija},  %(permanent address)}, \,\footnotemark[2], 
   Fr\'{e}d\'{e}ric Hecht\footnote{Sorbonne Universit\'{e}s, UPMC Univ Paris 06, UMR 7598, Laboratoire Jacques-Louis Lions, F-75005, %4 place Jussieu 75005,
                                   Paris, France}\,\footnote{CNRS, UMR 7598, Laboratoire Jacques-Louis Lions, F-75005, %4 place Jussieu 75005,
                                   Paris, France}\,\footnote{INRIA-Paris-Rocquencourt, EPC ***, Domaine de Voluceau, BP105, 78153 Le Chesnay Cedex}, 
   Slobodan \v Zumer\footnotemark[1]\,\footnote{In\v stitut Jo\v zef Stefan, Jamova 39, Ljubljana SI-1000, Slovenija}\\[30pt]
\end{large} 
\end{center}

%Small title:  A mesh adaptivity scheme and its efficiency

{\bf ABSTRACT}

\begin{small}
    This paper presents a 3D mesh adaptivity strategy on unstructured tetrahedral meshes %and implementation 
    by a posteriori error estimates based on metrics,
    studied on the case of a nonlinear finite element minimization scheme 
    for the Landau-de Gennes free energy functional of nematic liquid crystals. % with
    Newton's iteration for tensor fields is employed with steepest descent method possibly stepping in. 

    Aspects relating the driving of mesh adaptivity within the nonlinear scheme are considered.   
    The algorithmic performance is found to depend on at least two factors:  
    when to trigger each single mesh adaptation, and   
    the precision of the correlated remeshing.  
    Each factor is represented by a parameter, with its values possibly varying for every new mesh adaptation.  
    We empirically show that the time of the overall algorithm convergence can vary considerably 
    when different sequences of parameters  
    are used, thus posing a question about optimality.  

    The extensive testings and debugging done within this work on the simulation of systems 
    of nematic colloids substantially contributed to the upgrade of an open source finite element-oriented   
    programming language to its 3D meshing possibilities, as also to an outer 3D remeshing module.


  %Newton's method, 
  %steepest descent, 
  %functional minimization, 
  %phase transitions

  \noindent
  {\sl Corresponding author}: iztok.bajc@fmf.uni-lj.si \\
  {\sl E-mail adresses}: frederic.hecht@upmc.fr, slobodan.zumer@fmf.uni-lj.si.\\
  This work was partially supported by the European Commission 7OP Marie Curie ITN Hierarchy Project,  %, grant No. 
  and the Slovene Human Resources Development and Scholarship Fund (Ad Futura).  %, grant No.   
\end{small}

%%% 1 %%%%%%%%%%%%%%%%%%%%%%%%%%%%%%%%%%%%%%%%%%%%%%%%%%%%%%%%%%%%%%%%%%%%%%%%%
\section{Introduction}\label{s:introduction} 

    Effective minimization of functionals is an important topic in a variety of scientific tasks, 
    in which the increasingly powerful computational capabilities of the last decades had allowed 
    shifting from 2D systems to larger 3D ones. 
    \emph{Confined nematic colloids} with \emph{defects} in directional ordering fields \cite{b:science06} 
    are an example of such computational systems.     
    In fact, the \emph{Landau-de Gen\-nes free energy functional} \cite{b:deGennes1} for liquid crystals 
    is a representative nonlinear functional from theoretical physics, along with similar ones, 
    as for ex. the Gross-Pitajevski functional for Bose-Einstein condensates \cite{b:danaila}, 
    or the Ginzburg-Landau for superconductivity \cite{b:deGennes2}.  
    All of them are phenomenologically describing critical phenomena in condensed matter systems with 
    possible appearence of topological defects.  

    Advances in the computational science are relatively soon at hand for more classical computational fields, 
    e.g., fluid dynamics \cite{b:pironneau}, but usually not so readily used for theoretical physics porpouses, 
    at least in three dimensions.   
    Inter alia, the present paper tries to contribute also in this sense. 
    
    The Landau-de Gennes free energy functional \cite{b:deGennes1} is very well known in the realm    
    of liquid crystals science. Plenty of physical systems have already been simulated by its minimization 
    (for example \cite{b:science06,b:PRL07,b:PRE07}),  
    with a good qualitative agreement of such calculations with physical experiments, thus empirically 
    validating such approach. Also the mathematical task of well-posedness (existence and regularity) 
    of the minimizers for particular forms of the Landau-de Gennes functional has been successfully 
    analyzed \cite{b:gartland1}. 
    Finite elements were used in \cite{b:james06,b:confined}, 
    but without a truly systemic mesh adaptivity approach.
    The latter was employed in \cite{b:james09}, with an empirical mesh estimator, 
    upgrading the one used in a refining method \cite{b:fukuda} on a special symmetric case. 
    
    A lot of 3D simulations of \emph{nematic liquid crystals} (NLC) 
    employed the \emph{finite difference method} (FD). 
    In particular a set of codes, developed from methodologies 
    introduced in previous NLC hydrodynamics works (\cite{b:ds-2d} in 2D, and \cite{b:ds-3d} in 3D)
    has proved to be robust, and been successfully used 
    leading to some important theoretical results in the NLC field (see \cite{b:09} for an essential shorter r\'esum\'e). 

    \emph{Finite element methods} \cite{b:ciarlet} have the property that can considerably 
    decrease the number of degrees of freedom by use of \emph{unstructured tetrahedral meshes}.
    The latter can discretize the computational domain very flexibly, with (possibly larger) 
    variations of magnitude of the mesh tetrahedra. 
    Moreover, complicated surfaces can be modeled quite precisely with triangular surface meshes, 
    with usually well defined boundary conditions, and a broad set of theoretically well-founded error analyses. 
    On the other side, working with triangular and tetrahedral meshes implies an increased level of 
    complexity for their generation and manipulation, which for {\sl three-dimensional domains}
    is a field still reaching a complete operational maturity which could be 
    available to a broader public. 
    The present paper aims at contributing mostly to this point, in particular 
    to aspects concerning the driving of a nonlinear scheme along with mesh adaptivity.

    Concentrating on test examples of \emph{colloidal particles in confined nematic matrix} 
    (shortly, \emph{confined nematic colloids}), which present a challenging, almost singular 
    behaviour regarding mesh resolution requirements, the hereby presented scheme makes use 
    of the mesh adaptivity tool of \emph{metric mappings}, or shorter just \emph{metrics}. 
    These are representing a posteriori error estimates based on the Hessian of the solution(s),
    and are a still evolving \cite{b:hecht} subfield of mesh adaptivity.

    The overall scheme here used is programmed in \emph{FreeFem++}, a complete and free (open source) 
    C/C++ idiomatic programming language \cite{b:freefem} with powerful commands and data types dedicated 
    to the finite element method and its use for solution schemes of (systems of) partial differential equations
    and functional minimization. 
    The overall work for the present paper contributed to the development and smoothing of some of the meshing-related 
    parts used within it (with testing, debugging, and interacting with the modules' authors). 
        
    Summarizing, this paper will present a 3D mesh adaptivity strategy, based on isotropic metrics, 
    with a finite element algorithm, implementated in \emph{FreeFem++} on one processor,
    on the case of the minimization of the Landau-de Gennes free energy functional, modeling systems 
    of confined nematic colloids. The driving of mesh adaptivity coupled with the nonlinear scheme will 
    be found to be non-trivially dependent on parameters regarding tetrahedral meshes and metrics defined 
    on them.   
    The algorithmic behaviour with regard to two parameters will be particularly analyzed.
    Numerical experiments will show that the driving efficiency of the coupling of the nonlinear scheme 
    with mesh adaptivity depends at least on what are the stopping criteria at which a new mesh adaptation is triggered, 
    and to what precision each new mesh is rebuild. For the aims of this paper, sequences of different 
    parameter values have been choosed into fixed arrays, but for the future better solutions could be enivisaged. 
    The results and the correlations found along with the presented ideas and suggestions are supposed to be meaningful 
    for wider classes of nonlinear systems and physics typologies.

%%% 2 %%%%%%%%%%%%%%%%%%%%%%%%%%%%%%%%%%%%%%%%%%%%%%%%%%%%%%%%%%%%%%%%%%%%%%%%%
\section{Physical description of nematic liquid crystals} \label{s:physics}
    
\subsection{Nematic director and order parameters}
    
    \emph{Nematic liquid crystals} are an oily material that can flow like a liquid, but also 
    exhibit physical features (optical, for example), that are typical of crystals. 
    They are a \emph{mesophase}, i.e., more ordered than liquids, yet less ordered than crystals. 
    These properties are mostly due to the elongated, rod-like form of their molecules, that in an 
    appropriate temperature range (or under an applied electric/magnetic field) locally align into 
    a preferential axis, called the \emph{director} and denoted by a vector $\bf n$. 
    The degree of this alignment is described by another physical quantity, the \emph{scalar order parameter} $S$. 
    Both quantities are usually nonhomogeneous in space, thus formally represented by  a vector and a scalar field 
    ($\bf n ({\bf r})$ and $S ({\bf r})$), which can vary at each point of the nematic material. 

    Only its direction being important, the nematic director is defined as a unit vector (field),
    $|\bf n|=1$.
    Being the sense in which the nematic molecules are pointing (statistically) the same, 
    also the equivalence $\bf n \longleftrightarrow -\bf n$ must hold. 
    (Sometimes the set of possible vectors $\bf n$ in a certain point $\bf r$ of the nematic is described 
    mathematically with the equivalence class $S^2/\mathbb{Z}_2$, which figuratively means
    approximately a hemisphere of the Euclidean 2-dimensional sphere $S^2$ in $\mathbb{R}^3$,
    altough more precisely it is the real projective plane $\mathbb{R}P^2$.) 
    
    The possible values of the scalar order parameter $S$ range between -1/2 and 1. As the negative 
    values appear in situations not included here, we can concentrate our attention to the interval $\left[0,1\right]$.
    Here, termodynamically speaking, $S=1$ describes the \emph{ideal nematic phase}, in which all the molecules are 
    (would be) perfectly aligned, while the other extreme, $S=0$, describes the high-temperature \emph{isotropic phase}, 
    in which the kinetic energy of the molecules is so large that they are completely disordered, 
    like in a usual isotropic liquid. 
    For classical nematic materials intermediate bulk values of $S$ are the rule\footnote{Like, 
    e.g., $S\approx 0.53$ for pentylcyanobiphenyl (5CB), a well-known nematic material, extensively used 
    in physical experiments, with nematic phase at room temperature range, the properties of which (values 
    of physical constants) have also been used in the hereby simulations.}, representing 
    intermediate degrees of order.  
    
    The above description is not always enough to guarantee neither a truly correct physical 
    picture, nor computational stability.   
    Confined nematic systems with the inclusion of colloidal particles usually get frustrated, 
    leading to appearence of \emph{topological defects}. 
    The latter are regions (usually line-like), where the scalar order parameter drops to a lower value. 
    Some descriptions, as for ex. the above one with only director and scalar order parameter (in total only four scalar quantities), 
    can lead to singularities. 
    Thus, a second order tensor quantity must be introduced, that is, the \emph{tensor order parameter} $Q$, 
    related to the previous description by
    \begin{equation}\label{QfromSandP}
       Q=\frac{S}{2}(3 {\bf n} \otimes {\bf n} - I) + \frac{P}{2}( {\bf e}^{(1)} \otimes {\bf e}^{(2)} - {\bf e}^{(2)} \otimes {\bf e}^{(1)} ).
    \end{equation}
    Here, the greatest eigen value of $Q$ is the scalar order parameter $S$, and its correspondent eigen vector is the director $\bf n$. 
    The other two orthonormal eigen vectors are  ${\bf e}^{(1)}$, ${\bf e}^{(2)}$, and $P$ the \emph{biaxiality parameter}.
    When the latter may be negligible in some contexts, as, e.g., in setting boundary conditions, 
    the second term can be dropped, leading to an \emph{uniaxial approximation} model
    \begin{equation}\label{QfromS}
       Q({\bf r}) = \frac{S({\bf r})}{2}(3 {\bf n}({\bf r}) \otimes {\bf n}({\bf r}) - I).
    \end{equation}
    In both expressions $I$ is the $3 \times 3$ identity matrix and $\otimes$ the tensor product. As the hereby notation stresses,    %
    $Q ({\bf r})$ is a tensor \emph{field}, and thus its components $Q_{ij} ({\bf r})$ are scalar fields. 
    The tensor order parameter field is symmetric, $Q_{ij}=Q_{j\,i}$, and traceless, $Tr(Q)=0$, so it can be written as 
    
    \begin{equation}\label{tensorQ}
      Q =
    \left[
      \begin{array}{cccc}
      ~Q_{11} & ~~~~Q_{12} &     Q_{13} \\
         ~    & ~~~~Q_{22} &     Q_{23} \\
         ~    &    ~   & -Q_{11}-Q_{22} 
      \end{array} \right],
    \end{equation}
    where the non appearing lower off-diagonal components are meant to be equal to the corresponding symmetric upper ones. 
    As it can be seen, only five components of the tensor are needed to represent the whole tensor field.

\subsection{Landau-de Gennes model}  % and functional % theory

    From the \emph{Landau theory of phase transitions} \cite{b:landau,b:chaikin} it is known that appropriate 
    thermodynamical systems with an order parameter can be described in a suitable temperature range of a phase 
    transition by the {\sl Landau phenomenological} expansion of their free energy, under the condition that it 
    takes into account the symmetries of the system (i.e., the expansion must be invariant to them). 
    Well known applications of this theory are found for example in magnetic systems, in the temperature range 
    of the transition between the paramagnetic and ferromagnetic phase, or in superconductivity \cite{b:deGennes2}
    by the Ginzburg-Landau equations \cite{b:GinzburgLandau}, etc.   %  
    
    In our case, the \emph{Landau-de Gennes model} for liquid crystals will be employed,
    in which the \emph{Landau-de Gennes free energy functional} will have the form
    \begin{eqnarray}\label{Q}
       F(Q) &=& \int_\Omega  \left[ f_{e}(\nabla Q)  + f_{b} (Q) \right]dV + \int_{\Gamma_p} f_{s}(Q) dA.  
    \end{eqnarray}                                                             
    Here, the first integral comprises the volume contributions (elastic density $f_e$ and bulk $f_b$) to the 
    total nematic free energy in the interior of the (bounded) domain $\Omega$
    enclosing the space filled with nematic (thus {\sl without} colloidal particles, which are outside; $\Omega$ is thus a domain with holes). 
    
    The \emph{elastic energy} density $f_e$ can in general be constructed with three constants.
    Here we use a simplyfied but qualitatively still accurate version, employing the \emph{one-constant approximation}
    \begin{equation}\label{fE}
        f_e(\nabla Q) = \frac{1}{2} L\,|\n Q |^2  = \frac{1}{2} L \n Q_{ij} \cdot \n Q_{ij}
    \end{equation}
    with $L$ being the \emph{nematic elastic constant}. 
    The \emph{thermodynamic (}or \emph{bulk) energy} has the form
    \begin{eqnarray}\label{fB}
        f_b(Q) &=& \frac{1}{2} A~Tr(Q^2) + \frac{1}{3}  B~Tr(Q^3) + \frac{1}{4} C\,(Tr(Q^2))^2 \nonumber  \\
            &=& \frac{1}{2} A \, Q_{ij} Q_{ji} + \frac{1}{3}  B \,  Q_{ij} Q_{jk} Q_{ki} + \frac{1}{4} C \, (Q_{ij} Q_{ji})^2,
    \end{eqnarray}
    with $A,B,C$ being the \emph{material bulk constants}.  % [to add their values?]. 
    Here, as everywhere else in the text, Einstein double index summation is considered.
    
    The domain boundary $\p \Omega$ is splitted in two disjoint subsets, $\p \Omega = \Gamma_p \cup \Gamma_c$.  %, i.e.
    The first of the two, $\Gamma_p$, consists of the colloidal particles surfaces\footnote{Here only spherical, 
    but in general much more complicated shapes are possible.}   
    \, on which are defined the surface integrals, with penalty free energy density $f_s$, having the form of   % the  second
    the so-called \emph{Rapini-Papoular anchoring energy}
    \begin{equation}
        f_{s}(Q) = \frac{1}{2} W(Q_{ij} - Q_{ij}^0)(Q_{ij} - Q_{ij}^0), \label{fS}
    \end{equation}
    where the constant $W$ is the \emph{anchoring energy}, and $Q_{ij}^0$ the components of the reference tensor 
    order parameter field on the surface of the particles. 
    
    The second one, $\Gamma_c$, represents the computational cell walls, on which Dirichlet boundary conditions are employed,
    and which will be described in Section \ref{s:computations}.

\section{Free energy functional minimization \--- nematic structure calculation}  \label{s:minimization}  

    The nematic liquid crystal systems here considered are at constant temperature and   
    constant volume\footnote{Which also justifies the choice of the free energy $F$  
    as the appropriate thermodynamic potential.}. \,When such a system is physically let to evolve,
    its entropy grows driving the free energy potential to a minimum. 
    If the latter is global, the equilibrium is \emph{stable}, while if 
    just a local minimum has been reached, the structure is considered to be \emph{metastable}\footnote{In both cases 
    the nematic system is still fluctuating \cite{b:ziherl}, employing a \emph{statistical equilibrium}.}.  
    
    The Landau-de Gennes model is a static theory neglecting fluctuations. 
    Its free energy functional minima describe the equilibrium configurations of a nematic system. 
    Mathematically (computationally) this minimization can be achieved with finite elements in at least two ways.
    
    The first one uses the elementary and well known necessary condition \cite{b:gelfand} for a differentiable functional $F$ 
    to have a minimum for $Q=Q^*$, that is when its first variation vanishes, $ \delta F(Q^*) = 0$.  
    This in our case directly corresponds to the \emph{Euler-Lagrange equations} in weak form. %
    By solving this operator equation, i.e., by finding a numerical solution that sets it approximately to zero
    (here done by \emph{Newton's method} \cite{b:rheinboldt,b:gartland2}), one can achieve a minimum of the system. 
    In the second way a {\sl direct} minimization of the functional is performed, employing the \emph{steepest descent} method \cite{b:polak}. 
    In the present paper we use a hybrid technique, which starts in the first way, but possibly employs also the second one.

\subsection{Newton's method/Steepest descent} 
    
    The \emph{Newton's method} (or \emph{Newton-Raphson method}) is a second order approximate iterative method 
    for numerically solving (nonlinear) operator equations. Being the equation in our case  $ \delta F(Q) = F'(Q) \delta Q = 0$, 
    the iterative equation achieves the form
    \begin{equation}\label{e:NM1}
        F''(Q^{(i)})v^{(i)} \phi = -F'(Q^{(i)})\phi,
    \end{equation}    
    where $Q^{(i)}$ is the current solution at step $i$ (the last computed), and $F'(Q^{(i)})$ and $F''(Q^{(i)})$ 
    are correspondingly the \emph{first} and \emph{second variation} 
    of the functional $F$ in $Q^{(i)}$, $\phi$ are the test functions (written in place of $ \delta Q $), 
    and $v^{(i)}$ is the equation solution (i.e. the move, or increment field $v^{(i)} = Q^{(i+1)} - Q^{(i)}$ 
    at the $i$-th step between two successive iterations).  % (\ref{NM}) 

    Possible situations exist, in which Newton's method can fail. Its iteration sequence
    can diverge, for ex. at bifurcation points, i.e., when the second variation of the 
    free energy functional is singular, $det(F''(Q))=0$, or trap itself into a (quasi)periodic orbit  
    when in a neighbourhood of a saddle point. To overcome such problems, it seemed convenient 
    to introduce also a more stable method into which the algorithm could possibly switch  
    in these cases, i.e., in the neighbourhood of such problematic nematic configurations. 
    
    One of the oldest gradient algorithms for functional minimization is the \emph{steepest descent method}.
    Developed by Cauchy more than one century and half ago (altough for functions of several variables), 
    it is a robust iterative algorithm, suitable for such situations. 
    From an iterate $Q^{(i)}$ it proceeds by first calculating the (negative) gradient 
    $h^{(i)} = -\nabla F(Q^{(i)})$, then choosing a suitable parameter $\lambda$, and finally computing the 
    next iterate by
    The obtained trajectory somehow resembles the natural path of a droplet of water descending a hill
    under the force of gravity. 
    The main tasks to be accomplished during the steepest descent method are the calculation of the gradient,
    and the choice of the parameter $\lambda$, which was here done by the \emph{ exact line search}
    technique. As the Newton's algorithm, it must start in an initial configuration $Q^{(0)}$.

\subsection{Unstructured tetrahedral meshes and mesh adaptivity}  
    
    When numerically solving (systems of) partial differential equations a mesh is intrinsically  related to the 
    solution computed on it, and it can be said of \emph{good quality}, if leading to a \emph{good solution} \cite{b:frey}. 
    A computed solution is usually assumed to be such, if approximating the real solution with a \emph{low error}.  
    A general wish, aim, regarding meshes is to have the number of mesh vertices low as well. Assuming these definitions 
    and goals, a mesh can be considered \emph{optimal} \cite{b:frey}, if leading to a solution computed within a prescribed 
    error with the \emph{minimal} number of degrees of freedom. 
    
    Quantitatively, this can be obtained (and it is here done so, following \cite{b:alauzet-frey-cfd, b:dobrzynski-frey, b:alauzet-phd, b:dobrzynski-phd})
    by applying the \emph{equipartition} of the total error to all the mesh elements. After setting a relative interpolation error threshold, 
    the aim of the remeshing is to rebuild the tetrahedral mesh in such a way, that the interpolation error 
    would be everywhere, i.e. on each element, below it. 

    In general this can be best achieved by building mesh elements by varying their \emph{size},
    and varying also their \emph{shape} (i.e. edge lengths and angles between them) and \emph{orientation}. 
    The driving idea is that the size of the tetrahedra must get smaller in the regions of the computational domain,
    where the solution is spatially varying. 
    The more it varies, the smaller the elements must be in order to catch the solution shape correctly enough \---- below the prescribed error. 
    The motivation to locally vary also tetrahedra's shape and orientation is that if the solution 
    locally doesn't change much along a direction, than the tetrahedra in that direction can be more elongated.
    This implies the use of a much lower number of elements. Examples exist \cite{b:frey},
    where the number of degrees of freedom used (with P1 elements), has been decreased for
    ten times, compared to the same computations done with the isotropic approach. 

    A contribution of the present paper to similar ones regarding liquid crystals
    and exploiting similar features or methods, e.g. \cite{b:fukuda,b:james09,b:confined},  % at the best of our knowledge known to us,
    is the use of a systemic \emph{mesh adaptivity} approach on \emph{three-dimensional unstructured tetrahedral} meshes 
    based on \emph{isotropic metrics}.

    \subsubsection{Mesh generation}  
    
    The basic {\sl ideas} of unstructured mesh generation are quite similar irrespective 
    of the dimensionality of the space in which the mesh is built.
    In 1D, segments of different length are generated, while in 2D/3D triangles/tetrahedra 
    of different size, shape and orientation.  
    Nevertheless, the 3D case is technically much more difficult to implement than the previous two \cite{b:frey}. 
    
    Altough the implementation of mesh generation can be accomplished with an ample palette of approaches, 
    its basic underlying idea is substantially the same.  
    Regardless of the fact if we are using an \emph{advancing front technique}, or a \emph{Delaunay approach}, 
    a their combination, or something else \cite{b:frey}, we start with a closed surface mesh in 3D representing 
    the domain boundary.
    Then, using a criterion dependent of the technique used, we add vertices in its interior
    until the whole domain is tetrahedralized.   %al
    Such a tetrahedralization must simultaneously conform to both the \emph{domain geometry} 
    and the \emph{solution}. Thus, when choosing the position where a vertex should be added, 
    the information about both of them must be taken into account.

    \subsubsection{Metrics}  

    This can be achieved by employing \emph{metric mappings}, or simpler, \emph{metrics},
    with which it is possible to produce, via appropriate algorithms mentioned above,  
    unstructured meshes with tetrahedra of the locally desired size and orientation,
    possibly with large scale variations within the same domain $\Omega$.  % globally

    The main idea is that the usual (classical) Euclidean length in space
    \begin{eqnarray}
       d({\bf r}, {\bf r}')=||{\bf r} - {\bf r}'||_2 = \sqrt {< {\bf r} - {\bf r}', \,{\bf r} - {\bf r}'>}\, \label{euclidean}    
    \end{eqnarray}
    is changed by redefining the usual (Euclidean) scalar product $<\cdot\,, \cdot>$ in $\mathbb{R}^3$, 
    appearing in (\ref{euclidean}), with a new one, $<\cdot\,, \cdot>_M$, defined as  
    $$ <{\bf r},\,{\bf r}'>_M \,= < {\bf r},\,M {\bf r}'>,$$    
    with $M$ for being a constant symmetric positive definite matrix.  
    By leaving it vary over the computational domain $\Omega$,
    we obtain a $3 \times 3$ tensor field $\mathcal{M({\bf r})}$, called the \emph{metric tensor field}, or simply \emph{metric}. 
    With the domain $\Omega$ endowed with such a (Riemannian) structure, the theoretical distance $l_\mathcal{M}({\bf r}_1,{\bf r}_2)$ 
    between two points ${\bf r}_1, {\bf r}_2 \in \Omega$ now equals to
    \begin{eqnarray}
       l_\mathcal{M}({\bf r}_1,{\bf r}_2) = \int_0^1 \sqrt {< \gamma'(t),\, \mathcal{M}(\gamma(t)) \gamma'(t)>}\,dt,  \,    \label{e:varying}   
    \end{eqnarray}    
    where $\gamma$ is the shortest possible path (the geodesic) between the two points.
    For practical purposes, the average length 
    $ \bar l_\mathcal{M}({\bf r}_{12}) $ of an edge between two vertices ${\bf r}_1$, ${\bf r}_2$, described by  
    ${\bf r}_{12} = {\bf r}_2 - {\bf r}_1$, can be computed as
    \begin{eqnarray}
       \bar l_\mathcal{M}({\bf r}_{12}) = \int_0^1 \sqrt {< {\bf r}_{12},\, \mathcal{M}({\bf r}_1 + t {\bf r}_{12}) {\bf r}_{12}>}\,dt.  \,    \label{e:noneuclidean}   
    \end{eqnarray}
    As detailedly explained for ex. in \cite{b:frey, b:alauzet-frey-cfd, b:dobrzynski-frey, b:alauzet-phd, b:dobrzynski-phd}, 
    to have everything working correctly, $\mathcal{M({\bf r})}$ must be symmetric and positive definite. 
    This gives rise to \emph{anisotropic metrics}. 
    
    In the hereby calculations \emph{isotropic metrics} were used, 
    which means that the diagonalization of the tensor $\mathcal{M}$ in the local coordinate frame has 
    all of its three eigenvalues being equal. 
    The diagonal metric tensor is then equivalent to a spatial distribution of 
    the tetrahedral sides, i.e. a scalar field $h({\bf r})$.

    \noindent

\section{Computations}  \label{s:computations} 

For the computational examples we set and tested a code with calculations for the simplest case of confined 
colloidal nematic system, a colloidal particle immersed in confined nematic (i.e., the \emph{monomer}). 
The code has then been run for five different sizes of the system (all with 
the same length proportions), for three different types of convergence sequences 
(see explanation later), and for three different values of the computational parameter 
{\tt hmax} (see definition later on).

\begin{figure}[t]
  \scalebox{0.365}{\includegraphics{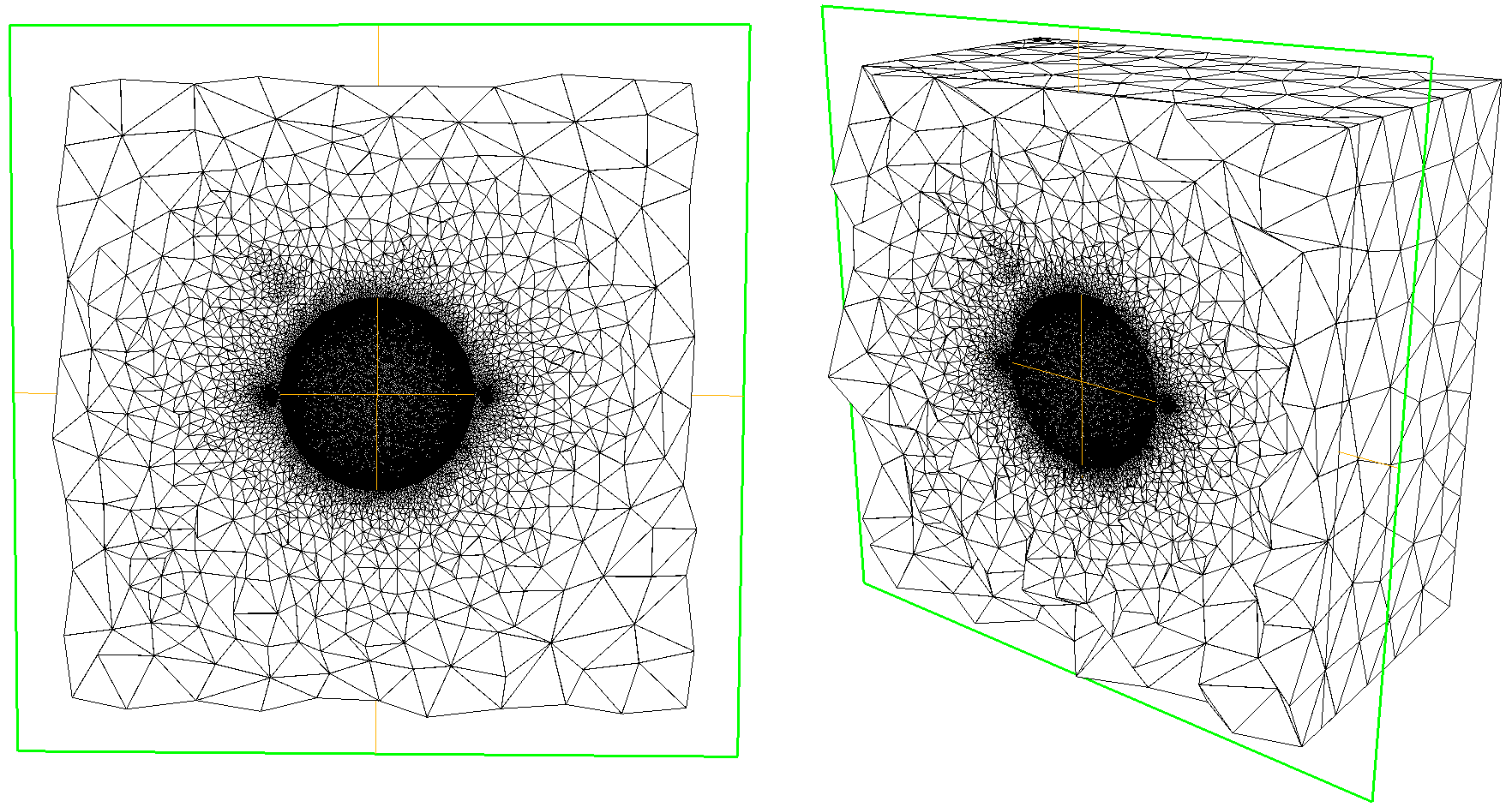}}
\vspace{-5mm}
\caption{\small Cross-section of the final tetrahedral mesh for one colloidal particle in confined nematic,
         obtained after the whole mesh adaptivity process: front perspective (leftx), and side perspective (right).
         Two cross-sections of the Saturn ring topological defect can be noticed symmetrically on 
         particle's sides, where the mesh is very refined. The sorrounding green square is 
         the cutting plane; tetrahedra at its intersection point out in hedgehog-style.
         Unstructured meshes can develop slight variations in density, e.g. near the particle,
         up on the left, which are mostly due to meshing technical reasons.} 
\label{f:Mesh-front+side} 
\end{figure}

\subsection{Experimental setting in physical laboratory}  

In the concrete experimental set-up in the physical laboratory (see for ex. \cite{b:science06}) 
the particles of spherical shape have a diameter of order of magnitude of a couple or some microns, 
and are usually made of silica, glass, or metal. They are immersed in a nematic liquid (here 5CB), 
contained inbetween two glass plates, distant some microns one from the other for a distance at least 
a couple of times the magnitude of the particle's diameter. 

The particles have \emph{homeotropic anchoring}, which means that their surface
is chemically treated with surfactant molecules attached perpendicular on it. 
Instead, the surfaces of the plates are treated mechanically (rubbed), in order 
to have horizontal anchoring with direction parallel to the sides of the cell. 
The nematic tends to align with the anchoring: at the sides of the cell parallel
to them, and on the surface of the particle perpendicularly to it.

\begin{figure}[t]
  \scalebox{0.365}{\includegraphics{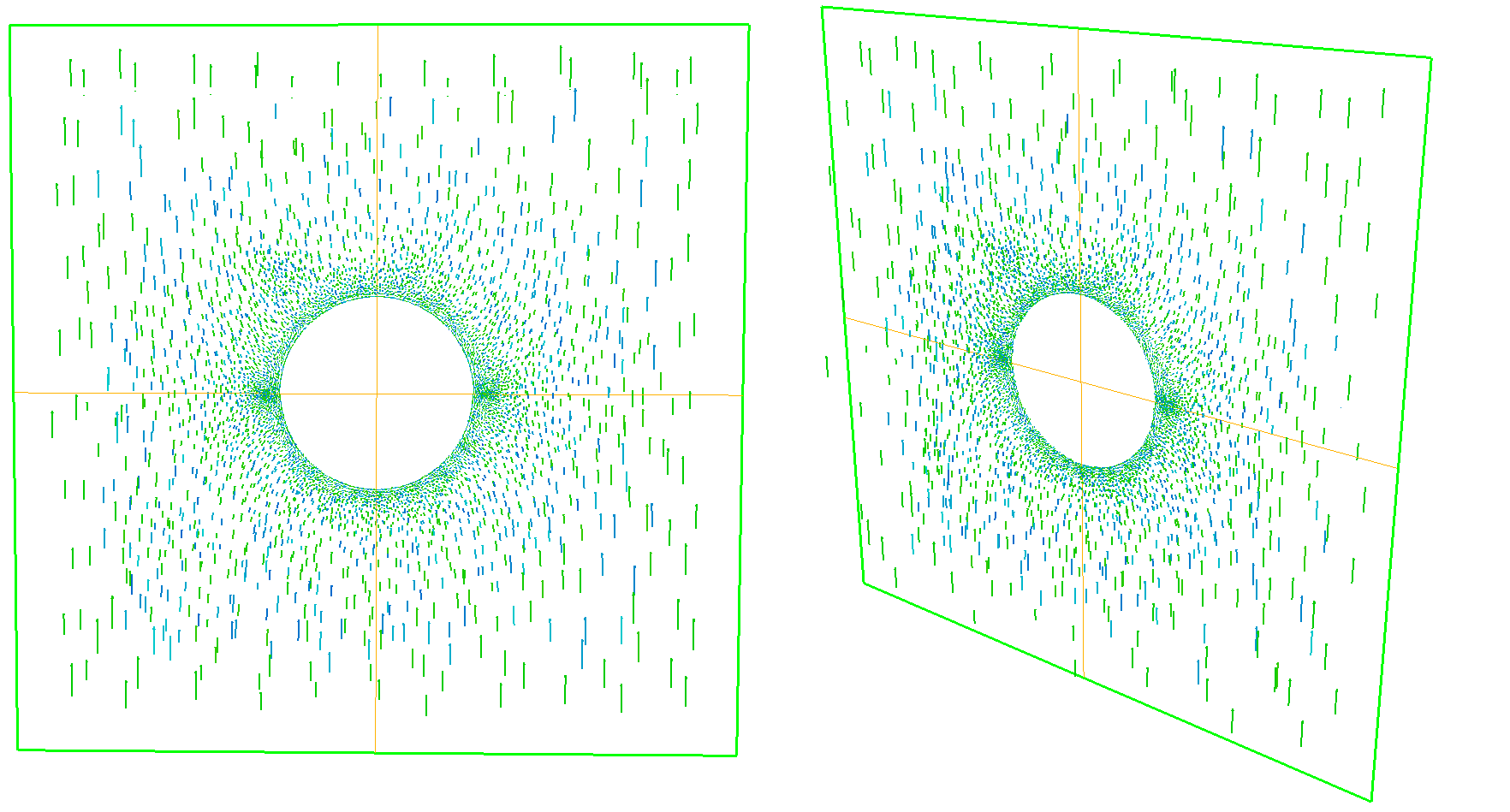}}
\vspace{-5mm}
\caption{\small Central cross-section of the computed nematic field around one colloidal particle
         correspondent to the final adapted mesh of Fig.\ref{f:Mesh-front+side}.   
         As before, two cross-sections of the Saturn ring topological defect can be noticed symmetrically on 
         the sides of the particle. 
         The visualization doesn't follow the standard LC literature for nematic fields.
         Theoretically, the length of each director ``vector'' (line) should be 
         always equal to one. Here, each length is proportional to the volume of the tetrahedron 
         on which it lies; consequenlty, the lines around the defect seem almost points.}  
\label{f:Dir-front+side-no-surf} 
\end{figure}

\subsection{Computational details} % set-up

The code has been written and tested first for the case of a spherical colloidal particle 
of diameter $2R = 1 \mu m$, posed in the center of a cubic cell with $d=2\mu m$, full of 
nematic with values of the material constants $L$, $A,B,C$ for the 5CB type
(for their values see for ex. \cite{b:science06,b:PRL07}). 
The boundary conditions matched the experimental ones described above. 
As at some sides the real (experimental) cell is very large (virtually infinite), and full of nematic, 
an approximation was made in the computations, putting at that sides, i.e., the walls of the computational cell, 
boundary conditions matching the behaviour of nematic at a longer distance.  

All the computations runned on one processor of a 64-bit desktop machine with 
Intel Core 2Quad CPU Q9550@2.83GHz$\times$4 processor, with 7.7GiB of RAM and 
the 64-bit {\tt Linux Ubuntu 12.04.4 LTS} operating system. 
The {\sl FreeFem++} version used was 3.30.
The main code in the remeshing process, {\tt mmg3d5ljll}, for the moment not part of the
standard {\sl FreeFem++} distribution yet, 
was cordially supplied by its authors Charles Dapogny, C\' ecile Dobrzynski, and Pascal Frey,
and called as an external module. 
All runs were reniced at their beginning to a nice value of -10, i.e., to a higher priority.

\subsubsection{Main scheme} \label{sss:main}

After being launched, the overall algorithm works in the following way (see Alg. 1, 
written in pseudo-code, below).  

First the initialization of the system is made. 
The initial mesh {\tt Th} is built, and the starting guess for {\tt Qh} set. 
Then, after the computation of the initial nematic structure into {\tt Qh}, with Newton iteration
(and possibly also steepest descent), the main loop is entered.
Here, at each iteration {\tt k} the current mesh is adapted into a new mesh, and a new nematic structure is computed on it.
This is looped for totally {\tt NbOfAdapt} times, which is a positive integer fixed by the length 
of the arrays of parameters {\tt tolAdapt} and {\tt errm}.   
Finally, the last computed mesh and nematic configuration on it are returned.\\ 

\noindent
{\bf Algorithm 1}
\vspace{-5mm}
\begin{verbatim}

// MAIN SCHEME:
Main(Sh, f, tolAdapt, errm)
{
  Initialize(Th, Qh; Sh, f);  // Initialization of Th and Qh.
  NbOfAdapt= length(tolAdapt);    // Total nb of adaptations.

  int k= 0;               // Adaptation index is initialized.
  Qh= Calculate_Nematic_Structure(Th, tolAdapt_k);  
  while (++k < NbOfAdapt)  {  
     Th= Adapt_Mesh(Th, Qh, errm_k);  
     Qh= Calculate_Nematic_Structure(Th, tolAdapt_k);  
  }   
  return Th, Qh;
}
\end{verbatim}  

%************************************
\begin{figure}[t]
   \scalebox{0.36} { \includegraphics{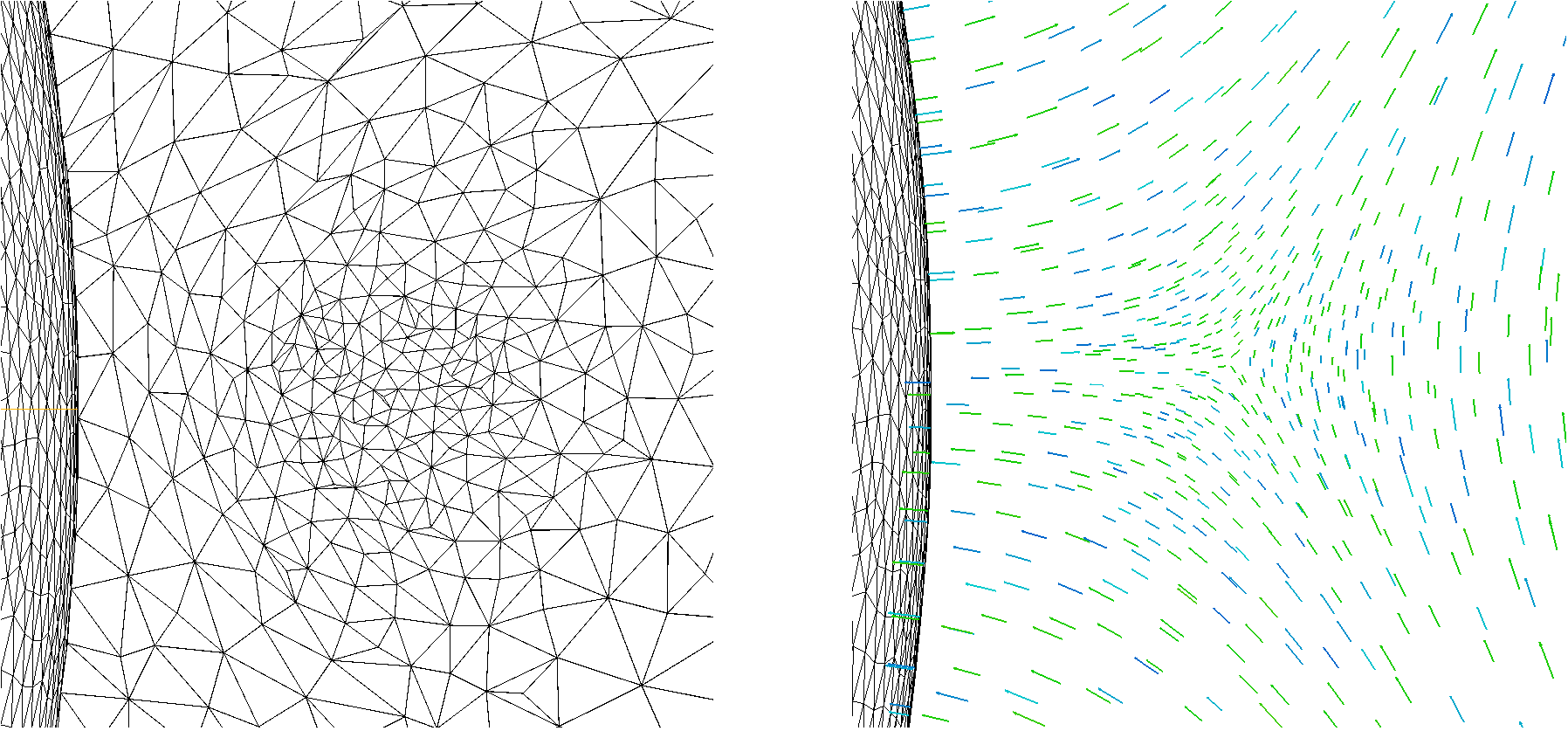} }
\vspace{-1mm}  
\caption{\small Zoomed enlargments of the mesh (left) and the corresponding nematic director field (right) from Figs. 
         \ref{f:Mesh-front+side} and \ref{f:Dir-front+side-no-surf} (both left)  
         around the (Saturn ring) topological defect.   
         In general, the volume (tetrahedral) mesh is more refined where the Hessian of the solution, or other appropriate functions, 
         is larger (e.g., around the defect), and/or where the domain geometry varies (e.g., near the sphere's surface).
         }  
\label{f:Defect-Mesh+Director-Zoom} 

\end{figure}

\subsubsection{Initialization: initial mesh and starting guess}  
    
    The surface mesh describing and enclosing the computational spatial domain was designed 
    within the {\sl FreeFem++} built-in functionalities, and then tetrahedrized with 
    {\tt TetGen} \cite{b:tetgen} as one of its inner modules. 
    More complicated surface meshes can be generated by {\tt Gmsh} \cite{b:gmsh}, 
    or other (free) mesh generators, and then imported into {\sl FreeFem++}.  
    
    A correct starting guess in this elementary case of a monomer was very simple, i.e., just the constant nematic configuration ${\bf n} = (0,0,1)$. 
    The initial tetrahedral mesh was set fine enough in the neighborhood of the particle, where stronger variations of the 
    nematic field and defects appear, and then linearly coarsened while approaching the cell walls, where the nematic 
    conformation changes no more.\\ 

\noindent
{\bf Algorithm 2}

\vspace{-5mm}
\begin{verbatim}

// INITIALIZATION:  
Initialize(Sh, f, Th, Qh)  
{
  Sh= Construct_Surface_Mesh();// Constructs main surface mesh.
  Th= Tetgen(Sh, fine_density);// Fine tetrahedrization.
  f=  Set_Initial_Mesh_Density(Sh);// Sets initial mesh density.
  Th= Tetgen(Sh, f);          // Tetrahedrizes with density f.
  Qh= Set_Starting_Guess(Th); // Starting guess is set.
  
  return Th, Qh;
}
\end{verbatim}

%************************************

\subsubsection{Newton iteration loop}  

    This is the core, or in any case one of the innest parts of the overall algorithm
    (the other one is the mesh adaptivity loop).  
    At each step $i$ of the loop, the Newton equation (\ref{e:NM1}) is solved. 
    First the finite element stiffness matrix and load vector are obtained
    by discretizing the variational equation (\ref{e:NM1}) on the fixed tetrahedral mesh $\mathcal{T}_k$,
    using P1 finite element basis functions. Being the sparse linear system symmetric 
    positive definite, it can thus be solved by the {\sl conjugate gradient method}
    (here with an $\varepsilon < 0.5  \cdot 10^{-7}$     
    relative error bound, and a rough preconditioner, dividing each line of the sparse matrix 
    with its largest element). This proved to be a good choice in this case,  
    being direct factorization methods, as also GMRES, impracticable, due to the large sizes of the systems.  
    The incremental solution $v^{(i)}$ of the sparse linear system is then 
    added to the current solution $Q^{(i)}$, obtaining $Q^{(i+1)}$, in which the  
    Newton step is again recomputed until the relative change of the functional value (of the system's free energy)  
    is lower than the tolerance  $tol_k$, or the maximal number of iterations is reached. 
    In the latter case the algorithm switches into the steepest descent method mood. 
    
    Alternatively, the normalized $L_2$-norm of the move (increment) $v^{(i)}$ 
    could be used as another (or concurrent) criterion. In our case was anyway being constantly monitored.

\subsubsection{Mesh adaptivity loop}   
Also each mesh adaptation itself is computed iteratively.     
First a new tetrahedral mesh variable {\tt Thx} is declared, 
which is then adapted several times during the loop. 
Its starting ``value'' is {\tt Th}, i.e., the last computed mesh before entering 
into the mesh adaptivity procedure.
Also a set of scalar fields {\tt scFields} is declared, with regard to which 
the metrics will be computed. 

Once the loop is started, at each new iteration a new finite element space {\tt Vhx}, 
based on the current mesh {\tt Thx}, is declared (which with {\sl FreeFem++} is done 
most easily and straightforwardly with just one short code line).   
An isotropic metric {\tt M} is then declared as a scalar field from this FE space, and 
computed with a call of {\tt mshmet}. 
One of the most important parameters of the latter call is {\tt errm\_k}, representing   
the largest possible relative error of the solution on each element, at the  {\tt k}-th iteration
(here ranging within a couple of percents, more precisely starting from 0.02 and ending with 0.01). 
The other parameter {\tt scFields} represent the scalar functions with regard to which the metric is computed. 
Initially these were only the five components $Q_{ij}$ of the tensor order parameter field, and the 
scalar order parameter $S$ (within the code written as {\tt Qh} and {\tt S}). 
After some experimentations it has been noticed and felt that also the 
inclusion of the (five) first variations of the free energy $\frac {\delta F} {\delta Q_{ij}}$ could 
make sense, and so they have been added to the list (as {\tt DF}). 

With this metric {\tt M} a new mesh is computed into {\tt Thx} by {\tt mmg3d5}.
The latter takes care that the mesh contains only tetrahedra with side lengths 
inbetween the (argument) parameters {\tt hmin} and {\tt hmax}, and also that 
the ratio between the side lengths of any two neighbouring tetrahedra does not 
exceed the prescribed \emph{mesh graduality} parameter value {\tt hgrad},
(here fixed to 2.00 throughout all the calculations). 

The loop ends when a condition characterizing some kind of convergence of the mesh 
(and/or the metric) is fullfilled. 
The condition must measure how much two subsequent meshes are close to each other. 
What we used here, was very simple, and most probably far from optimal, i.e., we 
used that the difference of the numbers of the vertices between two subsequent meshes 
in the loop does not exceed a certain number (here fixed to 300). 
Alternatively, another condition defined with the norm of the difference between two 
subsequent metrics could perhaps also be used, and would probably be more recommendable. 

In any case, the loop was set to stop at {\tt NAdaptIter} iterations (here fixed to 20).\\ 

\noindent
{\bf Algorithm 3}

\vspace{-5mm}
\begin{verbatim}

// MESH ADAPTATION: 
Adapt_Mesh(Th, Qh; hmax, errm_k)   // Other possible parameters: 
{                                  //  hmin, hgrad (here fixed).
   mesh3 Thx= Th; // Declares and initializes new mesh variable.
   scFields= {Qh, S, DF};  // Scalar fields for metric calculus.
   
   for (j=1; j<=NAdaptIter; ++j)  {
      fespace Vhx(Thx, P13d);          // Declares new FE space.
      Vhx M= mshmet(Thx, scFields, hmin, hmax, errm_k);//Metric.
      Thx= mmg3d5ljll(Thx, M, hmin, hmax, hgrad);  // Remeshing.
      if (meshes close enough)  break;   // Loop-exit condition.
   }
   return Th=Thx, Qh;  
}
\end{verbatim}
\vspace{-5mm}

\section{Numerical results}   \label{s:results}   

The present overall scheme is supposed to serve as a case for    
a wider class of nonlinear calculations. 
Many other, similar nonlinear problems, perhaps arising from different types of physics, 
are awaiting to be solved, or the methods for their solution waiting to be improved. 
A topic in this context that seemed quite important to our perception, and not so 
much treated until now, is {\sl how} to drive such a nonlinear algorithm in presence 
of mesh adaptivity.
The latter didn't appear a so clear task about which it could be straightforwardly 
possible to make definite statements. But with systematic examination by running 
many computations, some patterns could be noticed indicating a kind of behaviour.  

With the aid of collaborators who further developed the key codes {\tt mmg3d5} and {\tt mshmet}, 
i.e. their authors, we first made smoothly work the trinom composed by the two and {\sl FreeFem++}, 
the latter being the central programming language/software used, stron\-gly FEM-oriented, 
in which our main code was written. This meant smoothing out their functioning as single entities, 
as well as their interfacing/communication with {\sl FreeFem++}.

Plenty of preliminary tests were performed, in total several hundreds, may be thousand, 
each lasting from several hours to some days, on several cases of nematic colloidal systems. 
Apart from the monomer one, a lot of trials have been made also for the dimer, 
or for assemblies of several colloidal particles, i.e., for the so-called \emph{colloidal crystals}\footnote{In
these cases particular attention had to be brought to the setting of the 
starting guess.}, which could be two- or three-dimensional, as for ex. $2 \times 2$, or $2\times$3, or $2 \times 2 \times 2$, etc. 

\begin{figure}[t]
   \scalebox{0.465} { \includegraphics{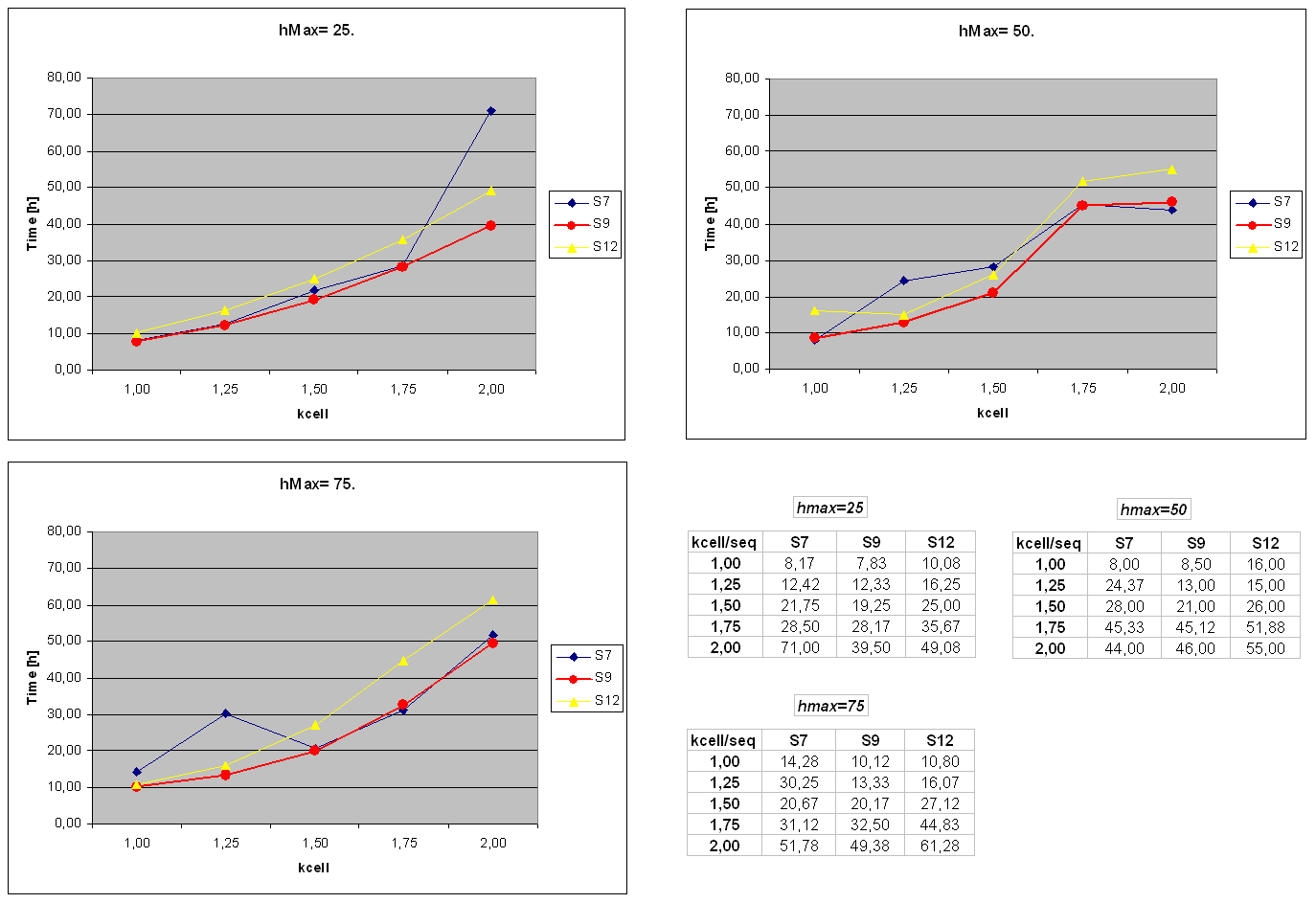} } 
\caption{\small Computation times for values of ${\tt hmax} = 25, 50, 75$. 
In all three cases the sequence S9 behaves better than the other two (almost everywhere).} 
\label{f:hMax25+50+75-Tables-C} 
\end{figure}

First it was recognized that the computations' behaviour of nonlinear finite elements based algorithms 
including mesh adaptivity is in general very parameter-dependent.
Changing the value of only one parameter can quite boldly modify the behaviour of entire sets of calculations.
This proved in the case of {\tt hmax}, the parameter representing the maximally allowed length of tetrahedral edges,
as it will be possible to notice further ahead, by comparing the computational results/measurements in the tables from Fig. \ref{f:hMax25+50+75-Tables-C}. 
 	
After these very extensive preliminary tests, and after the above mentioned computational trinom was set and working, 
we performed three sets of computations on the simplest of nematic colloidal cases, the monomer, for three values of {\tt hmax}. 
The latter indicated that the overall loop seems to be driven mostly by two factors.

\begin{table}[t]
\begin{tabular}[t]{|c||c|c||c|c||c|c|}
\hline
     Mesh    &  {\bf S7}  &         &  {\bf S9}  &         & {\bf S12}  &     \\
  adaptation &  {\tt tolAdapt}  &  {\tt errm}   &  {\tt tolAdapt}  &  {\tt errm}   &  {\tt tolAdapt}  &  {\tt errm}   \\
  \hline
       $0.$  &  0.5e-4  &  0.020  &  0.5e-4  &  0.020  &  0.5e-4  &  0.020  \\
       $1.$  &  0.5e-3  &  0.020  &  0.5e-3  &  0.020  &  0.5e-3  &  0.020  \\
       $2.$  &  0.5e-3  &  0.015  &  0.5e-3  &  0.015  &  0.5e-3  &  0.015  \\
       $3.$  &  0.5e-4  &  0.015  &  1.0e-4  &  0.020  &  1.0e-4  &  0.020  \\
       $4.$  &  0.5e-5  &  0.015  &  1.0e-4  &  0.015  &  1.0e-4  &  0.015  \\
       $5.$  &  0.5e-5  &  0.010  &  0.5e-4  &  0.015  &  0.5e-4  &  0.020  \\
       $6.$  &  1.0e-6  &  0.015  &  1.0e-5  &  0.015  &  0.5e-4  &  0.015  \\
 ${\bf 7}.$  &  1.0e-6  &  0.010  &  0.5e-5  &  0.015  &  1.0e-5  &  0.015  \\
       $8.$  &          &         &  0.5e-5  &  0.010  &  1.0e-5  &  0.015  \\
 ${\bf 9}.$  &          &         &  1.0e-6  &  0.010  &  0.5e-5  &  0.015  \\
       $10.$ &          &         &          &         &  0.5e-5  &  0.010  \\
       $11.$ &          &         &          &         &  1.0e-6  &  0.015  \\
 ${\bf 12}.$ &          &         &          &         &  1.0e-6  &  0.010  \\
\hline
\end{tabular}   %\\[3pt]
\caption{Three sequences (arrays) used in calculations.}  \label{t:sequences} 
\end{table}

The first one is {\sl when} (at what conditions) each new 
mesh adaptation is {\sl triggered}. This is determined by the threshold values of the free energy 
relative variations, and by how are they distributed throughout the nonlinear computation. 

The second factor influencing the algorithm's behaviour resulted to be {\sl how} the mesh adaptivity
is done, i.e., how the new mesh is rebuilt from the previous one at each mesh adaptation. 
This most strongly depends on the value of the solution error parameter, i.e. {\tt errm\_k}, 
appearing as argument in {\tt mshmet}. In fact, the call of the latter constructs the metric 
with which {\tt mmg3d5} then rebuilds the new mesh. 

On empirical basis of the hereby presented computations, we argue that a more general 
algorithm regulating both factors (and possible others, which weren't explicitly detected yet) 
should be a loop, or possibly several nested ones, with appropriate stopping conditions. 
We guess this could guarantee the larger flexibility needed for more general purposes. 
In fact we recognized (had the confirmation), as said before, that with finite elements based nonlinear algorithms 
with mesh adaptivity is in general not so easy to predict exactly how a nonlinear computation will 
behave, thus neither how much it will last before converging. 

Thus, to proceed by steps, we confined ourselves to arrange the threshold values in 
a {\sl fixed array} we called {\tt tolAdapt}, its constant length a priori determining 
how many times the mesh will be adapted, and its entries specifying at what thresholds. 
We set three such arrays, or sequences (see Table \ref{t:sequences}), calling them S7, S9, S12 
with their integer suffix being their length, and used each of them in a set of 
computations with varying size of the system, determined by its coefficient {\tt kcell}.

\begin{table}[t]
\begin{center}
\begin{tabular}[t]{|c|c|c|c|}
\hline
 {\bf kcell/seq.}  &  {\bf S7}  &  {\bf S9}  & {\bf S12}  \\
  \hline
  {\bf 1,00}	&  265285  &  261669  &  261552  \\
  \hline
  {\bf 1,25}	&  348853  &  346761  &  342469  \\
  \hline
  {\bf 1,50}	&  453226  &  437642  &  429681  \\
  \hline
  {\bf 1,75}	&  542873  &  534595  &  541236  \\
  \hline
  {\bf 2,00}	&  653378  &  625876  &  621104  \\
  \hline
\end{tabular}   %\\[3pt]
\caption{Numbers of vertices used in calculations for ${\tt hmax}=25$. 
Numbers for ${\tt hmax}=50$ and $75$ were similar, mostly slightly decreasing for 
a couple of percents with increasing values of {\tt hmax}.} \label{t:vertices} 
%\end{small}
\end{center}
\end{table}

Summarizing, what mostly drives the overall nonlinear algorithm is {\sl when} the mesh adaptivity
is triggered, and {\sl how} the respective new mesh is done. 
That is, at what {\sl free energy thresholds}, and within what {\sl errors}. 
An empirical proof of the fact, that it is not the same what strategy is brought into play, 
can be inferred from the tables in Fig. \ref{f:hMax25+50+75-Tables-C} , 
showing that computations with the se\-quen\-ce S9 were in almost 
all cases faster of those computed with the other two, S7 and S12, 
or in the worst case comparable \---- just slightly slower.

Regarding what properties the sequences of free energy threshold and mesh error values must have, 
it soon appeared quite evident that the values of the tolerances must be {\sl decreasing}.  %, from higher to lower ones. 
In fact, at the start of a single simulation run, the initially computed nematic conformations are usually % configurations
still quite far from the final (equilibrium) solution, i.e., the final nematic structure, 
and so the mesh adaptations must be more frequent.
Here, at each adapted mesh there's still no real need for convergence to a higher accuracy.
So the threshold values at the beginning of any sequence can be larger of those in the proceeding. 

Similarly, also the sequence of values of the solution error parameter {\tt errm} 
must tend to decrease, altough not necessarily completely monotonically.  
In reality, they must decrease at each (constant) threshold value.

Moreover, the present case suggested that the sequence of thresholds should be of intermediate 
length, as for ex. S9, i.e., neither too long, like S12, nor too short, like S7. 

Therefore, since typologies of physical systems and their sizes vary in general, and the parameter sequences should 
in general vary with them too, in both length and values composition, it seemingly should make sense that the use of sequences 
in fixed arrays could, as mentioned earlier, be changed in the future by the use of one or more {\sl loops}, possibly nested, 
satisfying suitable stopping conditions, that could drive the mesh adaptivity process optimally, or nearly so. 
\vspace{0.6cm}

\section{Conclusions and suggestions for the future}   \label{s:conclusions}

In this work a numerical method for functional minimization\footnote{As the minimization basically 
consists in resolving nonlinear systems of PDEs, the scheme can be used for them as well, thus regarded as more general.} 
of tensor fields on bounded, simply connected domains of Euclidean space $\mathbb R^3$  has been 
developed on the Landau-de Gennes case describing confined nematic colloidal systems.
Altough similar codes already exist, this finite elements based algorithm for the first time 
employs a systemic mesh adaptivity approach in 3D, with use of metrics (isotropic case). 

Anyhow, coupling the mesh adaptation process with the nonlinear scheme 
shows a strong parametric dependence. For the time being we solved it 
by a priori setting sequences of the driving parameters into fixed arrays.
Computations made for three different sequences, which were also of different length,
empirically demonstrated the parametric dependence and gave some insight into the 
process behaviour.  % into 

For a more general solution of this mesh adaptivity-driving task, that would 
be appropriate for a more ample class of nematic colloidal systems and other kind 
of physics problems, possibly dynamical, we imagine and would like to advocate 
the introduction of a special auxiliary algorithm, using for ex. several nested while-loops, 
which would be flexible enough for such purposes. 

When this and perhaps some other, more technical, questions
will be optimized, the here presented methodology could be 
assumed to be ready for more intensive calculations, 
aimed at systematic research in theoretical pyhsics. 

Extensions of the presented methodology could be envisaged also in 
directions of solving PDEs on more general manifolds\cite{b:iserles}, and/or the possible 
introduction of geometric integration \cite{b:hairer} for dynamical problems.

\section{Appendix}   \label{s:appendix}
    
    \subsection{First and second variation of $F$} 
        
        To implement equation (\ref{e:NM1}) into our code following \cite{b:gartland2},
        the first and second variation of the free energy functional 
        $F$ need to be calculated. We compute them analytically, the first one for both 
        Newton iteration and steepest descent, 
        and the second one just for Newton. 
        
        Before that, we must first of all solve the task of preserving  
        the traceless and simmetricity conditions of $Q$. This could have perhaps been done
        by the introduction of a special symmetric and traceless tensorial basis \cite{b:sonnet}, 
        which by construction preserves both conditions, as done for ex. in \cite{b:james06}. 
        Alternatively, we have opted to just apply the substitutions $Q_{33} = -Q_{11} -Q_{22}$ 
        and $Q_{j\,i} = Q_{ij}$ into the free energy expressions, after which the forms of the 
        free energy densities $f_e$, $f_b$ and $f_s$ depend only on the five components 
        $Q_{11}$, $Q_{22}$, $Q_{12}$, $Q_{13}$, $Q_{23}$. All the following calculations 
        have then been derived by taking into account only these components. 
        So the elastic free energy density becomes
        \begin{equation}\label{fE}
            f_{e}(\nabla Q) = L \left( \n Q_{ij} \cdot \n Q_{ij} + \n Q_{11} \cdot \n Q_{22}  \right),
        \end{equation}
        and the surface energy density
        \begin{equation}\label{fS}
            f_{s}(Q) = W \left[ (Q_{ij} - Q_{ij}^0) (Q_{ij} - Q_{ij}^0) + (Q_{11} - Q_{11}^0)(Q_{22} - Q_{22}^0)  \right].
        \end{equation}
        After some symbolic computer calculations, omitted here for brevity, also $f_b$ has been 
        transformed by substitutions into a polynomial of 4th degree in the actual five components $Q_{ij}$.
        
        Without digging too deeply into formalism, we will just assume that the previous  
        notation for the tensor field $Q$ will from now on mean the five-tuple $Q = (Q_{11}, Q_{22}$, $Q_{12}$, $Q_{13}$, $Q_{23})$,
        and similarly for all the other tensor field quantities, as $\delta Q$, $\varphi$, and $ v$. 
        For a formally exhaustive and more abstract treatment in a Sobolev space setting, the reader is referred to \cite{b:gartland1}.
        
        The {\sl first variation} of the Landau-de Gennes free energy $F$ (\ref{Q}) is  
        \begin{equation}\label{firstVarQij}
           \delta F(Q) = F'(Q) \phi =  \int_\Omega \left[ \frac{\p f_e}{\p \n Q_{ij}} \cdot \n  \phi_{ij} + \frac{\p f_b}{\p Q_{ij}} \phi_{ij} \right] dV + \int_{\Gamma_p} \frac{\p f_s}{\p Q_{ij}} \,  \phi_{ij}~ dA,
        \end{equation}
        where instead of $\delta Q_{ij}$ we already introduced the notation $\phi_{ij}$ for the test functions, 
        having well in mind that the pairs of indexes $ij$ have only the five couples of values defined above. 
        Variating again leads us to the {\sl second variation}  
        \begin{eqnarray}\label{secondVar}
           \delta^2 F(Q) = F''(Q) \phi v =  \int_\Omega \left[ \frac \p {\p \n Q_{kl}} \left( \frac{\p f_e}{\p \n Q_{ij}} \cdot \n \phi_{ij} \right) \cdot \n v_{kl} + \frac \p {\p Q_{kl}} \left( \frac{\p f_b}{\p Q_{ij}} \phi_{ij} \right) v_{kl} \right] dV \nonumber \\ 
                                                                                                                                                                + \int_{\Gamma_p} \frac \p {\p Q_{kl}} \left( \frac{\p f_s}{\p Q_{ij}}  \phi_{ij} \right) v_{kl}~ dA~.
        \end{eqnarray} 
        The terms of the first variation for the elastic part are now easily obtained as 
        \begin{eqnarray}
           \frac{\p f_e}{\p \n Q_{ij}} \cdot \n \phi_{ij} =  2L \left[ \n Q_{ij} \cdot \n \phi_{ij} + \frac 1 2 ( \n Q_{22} \cdot \n \phi_{11} + \n Q_{11} \cdot \n \phi_{22}) \right],   \nonumber
        \end{eqnarray}
        as also those for the second variation,
        \begin{eqnarray}
           \frac{\p }{\p \n Q_{kl}} \left( \frac{\p f_e}{\p \n Q_{ij}} \cdot \n \phi_{ij} \right) \cdot \n v_{kl} = 2L \left[ \n v_{ij} \cdot \n \phi_{ij} + \frac 1 2 ( \n v_{22} \cdot \n \phi_{11} + \n v_{11} \cdot \n \phi_{22}) \right],   \nonumber
        \end{eqnarray}
        where perhaps worth to be noted is the appearence of mixed terms. Similarly, the surface terms for the first variation are
        \begin{eqnarray}
           \frac{\p f_s}{\p Q_{ij}} \phi_{ij} =  2W \left[ (Q_{ij} - Q_{ij}^0) \phi_{ij} + \frac 1 2 \left( (Q_{22} - Q_{22}^0) \phi_{11} + (Q_{11} - Q_{11}^0) \phi_{22} \right) \right],   \nonumber
        \end{eqnarray}
        while those for the second read 
        \begin{eqnarray}
           \frac{\p }{\p Q_{kl}} \left( \frac{\p f_s}{\p Q_{ij}} \phi_{ij} \right) v_{kl} = 2W \left[ v_{ij} \phi_{ij} + \frac 1 2 ( v_{22} \phi_{11} + v_{11} \phi_{22}) \right],   \nonumber
        \end{eqnarray}
        where again similar mixed terms appear. The concrete calculations for both variations of the concrete $f_b(Q)$ 
        has been done with the help of the symbolic software \emph{Mathematica}.

  \subsection{Steepest descent} 
    \subsubsection{Gradient calculation} 
    
    The gradient, that we usually denote by $h$ (here with $h= -\nabla F(Q)$, following the notation of Polak \cite{b:polak}), 
    is an element of the Hilbert space, in which we are seeking the solution $Q^*$.
    For its calculation we use the Riesz theorem from basic functional analysis, which states that for each linear continuous 
    functional $G$, mapping from a Hilbert space $\mathcal H$ into $\mathbb R$, there exists exactly one element $h \in \mathcal H$, 
    such that the functional values $G(Q)$ is equal to the scalar product $<Q,\,h>$ for each element $Q$ from $\mathcal H$.  
    
    In our case the functional $G$ is the differential of the free energy functional $F$ in a configuration $Q$, i.e., $DF(Q)$. 
    Denoting now the gradient by $h$, and expanding it as $h = \sum_{i=1}^N h_i \phi_i$, i.e., 
    by the basis functions of the Hilbert space $\mathcal H$ 
    to which it belongs, we obtain
    \begin{equation}\label{scpr1}
        <Q,h> = \sum_{j=1}^N h_j <Q, \phi_j>.
    \end{equation}
    As we want this to hold for every $Q$, we set $Q=\phi_i$, for each $i$, getting
    \begin{equation}\label{scpr2}
        <\phi_i,h> = \sum_{j=1}^N h_j <\phi_i, \phi_j>, ~~~ i=1,\dots,N.  
    \end{equation}
    Being $<\phi_i,h>$ equal to the $i$-th gradient coefficient $h_i$, and $<\phi_i, \phi_j>$ to the $ij$-th element 
    of the Gram matrix, the calculation of the Hilbert space gradient is accomplished by first computing the 
    Gram matrix $K$, thus all the possible scalar products between the basis functions $\phi_i$, that is $K_{ij} \,= \,<\phi_{\, i}, \phi_j> $
    (here we note that the Gram matrix is sparse). 
    Besides, the negative of the differential (i.e. first variation) of $F$ is evaluated in the momentary configuration $Q$, 
    obtaining the right-hand side $d = \{-DF(Q)(\phi_i)\}_{i=1}^N$ of the linear system $ K h = d $. 
    By solving it\footnote{As before, we use conjugate gradients with relative tolerance $\epsilon=0.5\times 10^{-7}$.},~ 
    we obtain the gradient $h = \{-\nabla F(Q)(\phi_i)\}_{i=1}^N$.
    
    \subsubsection{Scalar product choice} 
        
        To implement this procedure, a choice of the scalar product must be made.
        %Here one has some freedom
        Following the structure of the Landau-de Gennes free energy, we define, 
        similarly as in \cite{b:danaila}, the scalar product as 
        \begin{equation}\label{SP1}
            <Q,P> := \int_\Omega \frac 1 2 \, L \, \n Q_{ij} \cdot \n P_{j\,i} + \frac 1 2 \, A \,Q_{ij} P_{j\,i} \,dV + \int_{\p \Omega} \frac{1}{2} W Q_{ij} \, P_{j\, i} \,dA,
        \end{equation}
        where Einstein summation is here for now employed over {\sl all} the indexes $i,j=1,2,3$. 
        The constant term $Q^0$ under the surface integral has been dropped 
        to preserve the definition scalar product property of $<Q,\,Q>$ vanishing only for $Q=0$, and
        the constants left to mantain appropriate proportions between the addends. 
        
        After applying the traceless condition $Q_{33}= -Q_{11} -Q_{22}$, and symmetricity $Q_{ij}= Q_{j\,i}$, we obtained
        \begin{eqnarray}\label{SP3}
            <Q,P> &=&  \int_\Omega L \, ( \n Q_{ij} \cdot \n P_{ij} + \frac 1 2 \, ( \n Q_{11} \cdot \n P_{22} +  \n Q_{22} \cdot \n P_{11} ) )  \nonumber \\ 
                  &+&  A \,( Q_{ij} P_{ij}  + \frac 1 2 \,( Q_{11}  P_{22} + Q_{22}  P_{11})) \, dV  \nonumber \\ 
                  &+& \int_{\p \Omega} W ( Q_{ij} \, P_{ij} + \frac 1 2 \, (Q_{11} P_{22} + Q_{22} P_{11}) ) \,dA
        \end{eqnarray}
        where, among the Einstein summation through only five index pairs, 
        additional mixed terms in the index pairs 11 and 22 appear. For {\sl exactly} traceless (and symmetric) tensor fields 
        this {\sl is} a scalar product. But for tensor fields which are numerically non-exactly traceless, it is no longer such, 
        lacking again the property that $<Q,Q>$ has to vanish only when $Q=0$. 
        Thus, after some experiments the mixed terms has been dropped, finally leaving 
        \begin{eqnarray}\label{SP3}
            <Q,P> &=&  \int_\Omega \left( L \, \n Q_{ij} \cdot \n P_{ij} +  |A| \, Q_{ij} P_{ij} \right)\, dV  + \int_{\p \Omega} W Q_{ij} \, P_{ij}  \,dA,
        \end{eqnarray}
        where the absolute value brackets has been added to the constant $A$, otherwise the product could sometimes be negative,
        and thus obviously contraddicting the nonnegativity condition of the scalar product.

    \subsubsection{Exact line search} 
    
        At each steepest descent iteration the calculated gradient gives only the direction of the maximal descent,
        but lets unsolved how much one should move in this direction.
        Thus, the iteration step must include also the choice of a proper coefficient $\lambda \geq 0$. This 
        can be crucial for the convergence itself, as for the time dependence of the iteration. 
        The optimal choice for $\lambda$ is the solution of the minimization problem
        \begin{equation}\label{lineSearchDef}
           \lambda^* = \arg \min_{\lambda} \{F(Q + \lambda h)\},
        \end{equation}
        which is called \emph{exact line search}. This can seldom be too expensive, 
        thus leading to a preference for approximative methods as for example the Armijo method \cite{b:polak}.
        But in the present case it leads to a not too complex or expensive situation. 
        For the Landau-de Gennes functional the problem (\ref{lineSearchDef}) means 
        \begin{eqnarray}\label{lineSearchExt}
           F(Q + \lambda h) &=& \int_\Omega  \left[ f_{e}(\nabla Q + \lambda \nabla h)  + f_{b} (Q + \lambda h) \right]dV + \int_{\Gamma_p} f_{s}(Q + \lambda h) dA,
        \end{eqnarray} 
        which can be quite easily expanded and collected with regard to powers of $\lambda$, here 
        done once with {\sl Mathematica}, and then transcribed into the {\sl FreeFem++} code. 
        After obtaining the coefficient terms by integration (here done within the code), 
        one in fact gets a polynomial  of fourth order in dependence of  $\lambda$: 
        $$p(\lambda) = a_0 + a_1 \lambda + a_2 \lambda^2 + a_3 \lambda^3 + a_4 \lambda^4. $$ 
        Extremal values are found when $p'(\lambda)=0$, that is, when
        $$p'(\lambda) = a_1 + 2 a_2 \lambda + 3 a_3 \lambda^2 + 4 a_4 \lambda^3 = 0, $$ 
        the three zeros of which are found with a GSL numerical procedure \cite{b:GSL}. 
        The minimal root between them is taken as the optimal $\lambda^*$, after a verification of the 
        positiveness of $p''$ in it as the minimum condition. 
        During concrete computations $\lambda^*$ usually ranged around values between 0.01 and 0.3, while  the other 
        two roots were almost always pairs of complex conjugated zeros, thus not feasible candidates.\\[3pt]

{\noindent \bf Acknowledgments.}
   Miha Ravnik, Daniel Sven\v sek, and George Mejak (University of Ljubljana)  % from FMF,
   are acknowledged for useful discussions, and Pierre-Henri Tournier (UPMC, Paris)
   for help with debugging of the {\sl FreeFem++} related internal and external modules.

%%%%%%%%%%%%%%%%%%%%%%%%%%%%%%%%%%%%%%%%%%%%%%%%%%%%%%%%%%%%%%%%%%%%%%%%%%%%%%%

\end{document}